\documentclass[twocolumn,aps,pre,showpacs]{revtex4}
\usepackage{epsfig}

\begin{document}

\title{
Numerical study of domain coarsening in anisotropic stripe patterns
}
\author{Denis Boyer}

\email{boyer@fisica.unam.mx}

\affiliation{ Instituto de F\'\i sica, Universidad
Nacional Aut\'onoma de M\'exico, Apartado Postal 20-364, 
01000 Mexico City, Mexico
}
\date{\today}

\begin{abstract} 
We study the coarsening of two-dimensional smectic 
polycrystals characterized by grains of oblique stripes with only
two possible orientations. For this purpose,
an anisotropic Swift-Hohenberg equation is solved. 
For quenches close enough to the onset of stripe formation,
the average domain size increases with time as $t^{1/2}$.
Further from onset, anisotropic pinning forces similar to
Peierls stresses in solid crystals slow down defects, and growth becomes
anisotropic. 
In a wide range of quench depths, dislocation arrays remain mobile
and dislocation density roughly decays as $t^{-1/3}$,
while chevron boundaries are totally pinned. 
We discuss some agreements and disagreements found with recent experimental 
results on the coarsening of anisotropic electroconvection patterns. 
\end{abstract}
\maketitle

\section{Introduction}

Coarsening occurs when a system is rapidly quenched below
a transition point into a phase with broken symmetries.
The spatio-temporal evolution following a quench
is relatively well known when the broken symmetry phase is characterized by 
a local order parameter that is spatially uniform ({\it e.g.} 
the local magnetization of a ferromagnetic phase) 
\cite{re:gunton83,re:bray94}.
The growth of spatial correlations, driven by domain growth or 
the annihilation of topological defects, usually obeys a dynamical 
scaling relation and the correlation length, or \lq\lq domain size" $R$,
grows as a power law of time with a well defined exponent. 
Classification schemes have been established for the main different cases 
\cite{re:hohenberg77,re:bray94}.

The situation is much less understood for phases characterized by a
local order parameter that is spatially modulated. Systems
forming periodic patterns (stripes, hexagons) with a well defined periodicity
can be observed in numerous physical systems, such as
Rayleigh-B\'enard convection, diblock-copolymer melts, magnetic materials, 
or Turing reaction-diffusion systems \cite{re:cross93}. 
After a quench into a stripe phase, two dimensional configurations
are composed of many domains differently oriented, including grain boundaries, 
dislocations and disclinations.
Numerical \cite{re:elder92,re:elder92b,re:cross95a,re:hou97,re:shiwa96,
re:christensen98,re:boyer01b,re:boyer02a,re:qian03}
as well as experimental 
\cite{re:purvis01,re:kamaga03,re:harrison00b,re:harrison02} studies 
have established that it is difficult, if not impossible, 
to reduce the ordering dynamics of stripes to one of the class known 
for uniform phases. There is still some debate
regarding the growth mechanisms, the value of the growth exponent, 
whether dynamical scaling holds or not, or whether the system may involve 
various characteristic length scales growing with different exponents.
In contrast with uniform phases, 
the coarsening rates depend significantly on the quench depth. Far away from 
the bifurcation threshold of stripe formation (large quenches), 
numerical solutions of the Swift-Hohenberg equation show that coarsening stops 
at large time, {\it i.e.} the system  remain frozen in macroscopically 
disordered configurations \cite{re:hou97,re:boyer02a}.
Based on an analysis of the law of motion of a grain boundary through
curved stripes, it was recently proposed that a single growth exponent 
could be introduced, but for vanishingly small 
quenches only \cite{re:boyer01b,re:boyer02a}. From dimensional arguments, 
a $R\sim t^{1/3}$ growth law was derived in that regime, in good agreement 
with numerical results at small quenches 
\cite{re:boyer01b,re:boyer02a,re:qian03}. 
The freezing observed at finite quenches was attributed to the presence of a 
periodic pinning potential (generated by the pattern itself) acting 
on grain boundaries.

 In the present paper, we consider a closely related problem where similar
questions remain open, and that has not been investigated 
numerically so far: the coarsening of {\it anisotropic} stripe patterns. 
Oblique rolls making only two possible angles
($\theta$ or $-\theta$, fixed) with respect to a particular axis
can be observed in electroconvection of nematic liquid crystals 
\cite{re:cross93}. Studying the ordering dynamics on this system is motivated
by various reasons. First, one can intend a comparison with available 
experimental data, since coarsening experiments have been recently conducted 
in electroconvection \cite{re:purvis01,re:kamaga03}. 
Second, the polycrystalline structures of oblique stripes
have a relatively simpler geometry than those of 
isotropic stripes: the constraint of the fixed angle
prevents the formation of disclinations. Therefore, the topological 
defects are essentially dislocations and (chevron) grain boundaries separating 
domains differently oriented. This situation can be seen 
as a smectic analogue of the structures formed by grains in 
polycrystalline solids, where 
disclinations are also absent \cite{re:chaikin95}. 

 We consider in the following an extension of the Swift-Hohenberg equation 
for oblique stripes in two spatial dimensions. 
This model, proposed by Pesch and Kramer for
describing electroconvection \cite{re:pesch86}, is recalled in 
Section \ref{sec:model}. In Section \ref{lowq}, we investigate
{\it small} quenches: the numerical results show that coarsening 
is driven by surface tension and a growth law $t^{1/2}$ is observed
for various characteristic length scales, like in 
Model A \cite{re:hohenberg77}. The results qualitatively change at larger
quenches (Section \ref{sec:pinning}): the characteristic length scales
associated with dislocations and chevron boundaries start to evolve 
differently, and the associated effective 
growth exponents progressively decrease as the quench depth increases. 
However, the effective exponent of the dislocation density 
remains fairly constant for a relatively wide range of quench depths. 
This feature can be explained by the fact that 
dislocations have a much lower pinning potential than chevron boundaries. 
We qualitatively justify this feature from weakly nonlinear 
analysis arguments. The dislocation exponent is close to the 
value of $1/3$ in that intermediate range, in agreement with the value 
measured in recent experiments \cite{re:kamaga03}. Some conclusions 
are presented in Section \ref{sec:concl}.

\section{Model equation}\label{sec:model}

Electroconvection in nematic liquid crystals is a paradigm of anisotropic 
pattern formation \cite{re:kai89}.  
If a nematics is placed between two glass plates properly treated, 
its director can be aligned along a preferential direction, say the $x$-axis. 
When an external a.c. electric field is applied in the direction normal 
to the plane, periodic rolls appear above a threshold. 
As the voltage is increased (the frequency being fixed in some proper range), 
bifurcations to various phases can be observed: \lq\lq normal" rolls, 
with a wave vector directed along the $x$-axis, usually appear first. 
This phase can be followed by a transition to \lq\lq oblique" rolls, 
of interest here, characterized by a wavevector with two possible 
orientations with respect to the $x$-axis, $\theta$ and $-\theta$. 

Although the theoretical understanding of electroconvection patterns based 
on constitutive equations is still incomplete, some
nonlinear models that rely on equations for a local order parameter
and on symmetry arguments have been proposed. 
Some time ago, Pesch and Kramer introduced an anisotropic
model \cite{re:pesch86} that exhibits a transition from normal to 
oblique rolls:
\begin{equation}\label{pk}
\frac{\partial \psi}{\partial t}=r\psi-\zeta^4(\Delta+k_0^2)^2\psi
-\frac{c}{k_0^4}\partial_y^4\psi
+\frac{2\eta}{k_0^4}\partial_x^2\partial_y^2\psi-\psi^3,
\end{equation}
with $\partial_{x(y)}=\partial/\partial x(y)$.
In equation (\ref{pk}), $\psi(\vec{x},t)$ is a local dimensionless order 
parameter, interpreted as 
a small lateral elastic displacement; $c$ and $\eta$ are dimensionless
constants modeling the loading forces and anisotropic bending constants;
$k_0$ is the wavenumber of the base periodic pattern, and $\zeta$ a
\lq\lq coherence" length that will be set to $1/k_0$ for simplicity here.
The dimensionless parameter $r$ is chosen as the main control parameter.
For $c=\eta=0$ the above equation reduces to the well known
Swift-Hohenberg model of Rayleigh-B\'enard convection
($r$ is the reduced Rayleigh number in that case).
The model (\ref{pk}) derives from a Liapunov \lq\lq free-energy" 
functional. It can be recast as
\begin{equation}\label{pot}
\frac{\partial \psi}{\partial t}=-\frac{\delta F}{\delta \psi}
\end{equation}
with
\begin{eqnarray}\label{liap}
F&=&\frac{1}{2k_0^4}\int d\vec{r}\left[k_0^4(-r\psi^2+\psi^4/2)
+\psi(k_0^2+\Delta)^2\psi\right.\nonumber\\
&\ &\left. -2\eta(\partial_x\partial_y\psi)^2
+c (\partial_y^2\psi)^2\right].
\end{eqnarray}
The functional $F$ monotonically decreases with time, $dF/dt\le 0$.

Linear stability analysis of Eq.(\ref{pk}) around 
the state $\psi(\vec{x},t)=0$ shows that two modes of finite wavenumber 
$\vec{k}=p\hat{x}+q\hat{y}$ ($\hat{x}$ and $\hat{y}$ are unitary vectors) 
become marginally unstable when the control
parameter $r$ increases and crosses some threshold values $r_c^{{\rm (o)}}$
and $r_c^{{\rm (n)}}$:
\begin{eqnarray}
&\ &r_c^{{\rm (o)}}=\frac{-\eta^2}{c+2\eta-\eta^2}<0,\  
\{\ p_c^2=\frac{k_0^2(c+\eta)}{c+2\eta-\eta^2},\nonumber\\ 
&\ &\quad\quad\quad\quad q_c^2=\frac{k_0^2\eta}{c+2\eta-\eta^2}\ \} 
\ \ {\rm (oblique\ rolls)} \label{oblique}\\
&\ &r_c^{{\rm (n)}}=0,\ \ \{\ p_c^2=k_0^2,\ q_c^2=0\ \}\ \ 
{\rm (normal\ rolls)} \label{normal}
\end{eqnarray}
In the above relations, we have considered the case 
$c>0$, the condition for which the instability at $r_c=0$ is to normal rolls
($p_c=0$). In the oblique phase, rolls make an angle $\theta=\pm 
\arctan[\sqrt{\eta/(c+\eta)}]$ with the $y$-axis. As obvious from 
Eq.(\ref{oblique}),
oblique rolls can only be observed for $\eta>0$. Hence, when $\eta$ 
is tuned from negative to positive values, a transition form normal 
to oblique rolls can occur. In order to study the coarsening
of oblique rolls, we will chose $c>0$ and $\eta>0$ in the following. 

Oblique rolls have the lowest threshold value ($r_c^{(o)}<0$) and we 
re-note $r_c\equiv r_c^{(o)}$ for simplicity. 
As already noted by Pesch and Kramer, the structures that are
likely to be observed for $r$ in the range $[r_c,0]$ may not be oblique rolls, 
but more complicated, nonlinear structures (\lq\lq undulated" rolls) 
that are not of interest here. When numerically solving Eq.(\ref{pk}) with 
random initial conditions and $r_c\le r\le 0$,
we actually observed that oblique rolls never appeared. On the other hand, 
configurations of oblique rolls very similar to those
observed experimentally 
\cite{re:purvis01,re:kamaga03} 
are always observed when setting
$r>0$ instead. Normal rolls were never observed in the runs presented 
in the following Sections. 
This is consistent with the weakly nonlinear analysis of Eq.(\ref{pk}) 
that predicts that oblique rolls have a 
lower free-energy $F$ \cite{re:pesch86}. 
Therefore,  we define the quench depth as
\begin{equation}\label{defqu}
\epsilon=r-r_c, 
\end{equation}
with $r_c$ given by Eq.(\ref{oblique}). We always chose
$\epsilon$ larger than $|r_c|$  ($r>0$).

\section{Coarsening kinetics near onset ($\epsilon\ll 1$)}\label{lowq}

\begin{figure}
\epsfig{figure=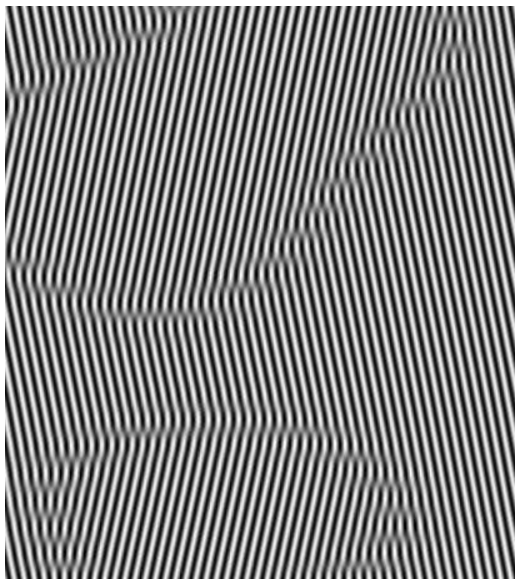,width=3.in,angle=90}
\caption{\label{t1200} Local order parameter in gray scale (detail), $c=12$,
$\eta=0.5$, $\epsilon=0.0372$, t=1200, obtained from random initial 
conditions.}
\end{figure}
\begin{figure}
\epsfig{figure=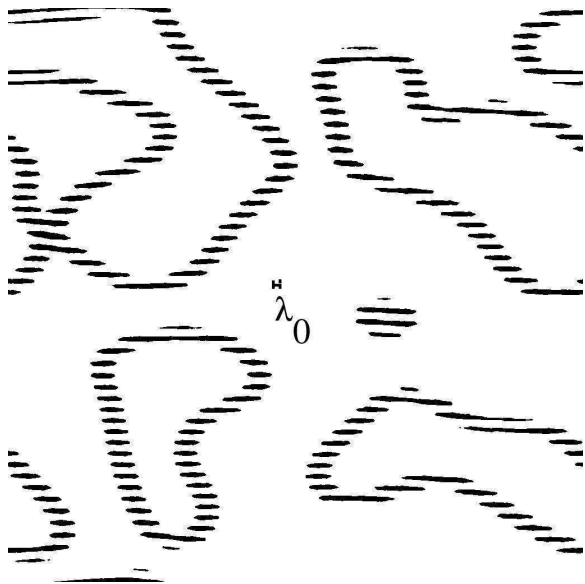,width=3.in,angle=0}
\caption{\label{t1200def} Defects (marked in black) of a configuration 
with same parameters as in Fig.\ref{t1200}, at a larger scale.}
\end{figure}

We numerically solve Eq.(\ref{pk}) by using a pseudo-spectral method
and a time integration procedure whose
descriptions can be found in Ref.\cite{re:cross94}.
The space is discretized on a square lattice of $1024^2$ nodes
with a lattice size $\Delta x$ set to unity. The base period 
$\lambda_0\equiv 2\pi/k_0$ of the pattern is fixed to $8\Delta x$.
The time integration scheme is stable for relatively large value
of the time step, which is fixed to $0.5$ in dimensionless time units.
The initial condition for $\psi$ is a random field with Gaussian 
distribution, of zero mean and variance $\sqrt{\epsilon/3}$.

Figure \ref{t1200} displays in gray scale the order parameter $\psi$ at time 
$t=1200$ time units, for a run with $c=12$ and $\eta=0.5$ (the angle of 
the rolls with the vertical axis is $11.31^{\circ}$). The quench depth is
{\it small}, and has been set to $\epsilon=1.9|r_c|\simeq 0.0372$.
The configuration is that of a smectic polycrystal:
Most of the defects present are grain boundaries separating
zig and zag rolls, and few isolated dislocations can 
be observed. Due to the asymmetry of the problem, two kind of boundaries
can be roughly distinguished: \lq\lq horizontal" chevron boundaries where the 
roll orientation changes rather smoothly from one grain to the other,
and \lq\lq vertical" (or inclined) boundaries, that are made of dense
arrays of dislocations. This distinction is not very sharp, as one
go continuously from one situation to the other, corresponding to
boundaries of \lq\lq low" and \lq\lq high" 
dislocation density, respectively.
The defect field shown in Figure \ref{t1200def} is obtained from
Fig.\ref{t1200} by using a Fourier filtering procedure.
The dark areas correspond to defected regions. 
Dislocations tend to be distributed along string-like structures,
like in crystals.

At very large times, grain boundaries are weakly curved and
isolated dislocations lying inside a grain are relatively rare.
These observations agree with recent electroconvection 
experiments \cite{re:kamaga03}, where
a mechanism for the formation of isolated dislocations was identified:
A shrinking bubble can be roughly pictured as delimited
by two vertical and two horizontal grain boundaries. In some cases, the 
two vertical boundaries are not composed by the same number of 
dislocations, therefore, some dislocations can not annihilate with 
others of opposite Burgers vector when the bubble shrinks.
However, this situation occurs rarely.

\begin{figure}

\vspace{-2.5cm}
\hspace{-2.5cm} \epsfig{figure=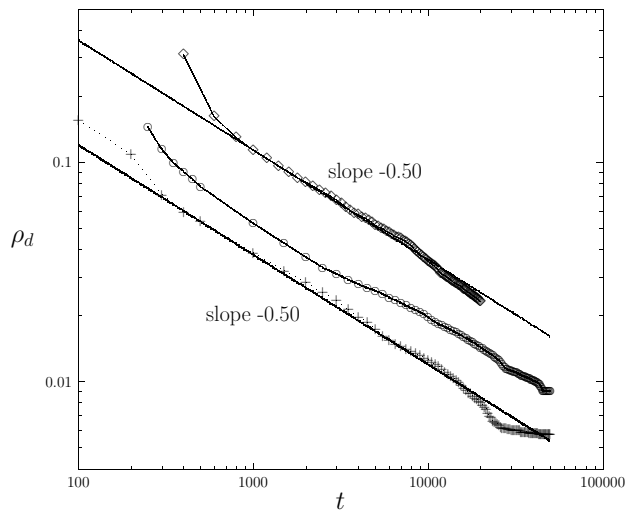,width=4.in}
\vspace{-5.5cm}
\caption{\label{gb} Defect density as a function of time. 
From bottom to top.
($+$) symbols : $c=12$, $\eta=0.5$ ($\theta=11.31^{\rm o}$), 
$\epsilon=1.9|r_c|= 0.0372$.
($\diamondsuit$) symbols : $c=6$, $\eta=0.25$ ($\theta=11.31^{\rm o}$),
$\epsilon=1.9|r_c|= 0.0184$.
($\bigcirc$) symbols: $c=3$, $\eta=0.25$ ($\theta=15.50^{\rm o}$) , 
$\epsilon=1.9|r_c|= 0.0345$.
Average are performed over 13 runs in each case. Solid lines
are guides to the eye.}
\end{figure}

\begin{figure}

\vspace{-2.5cm}
\hspace{-2.5cm} \epsfig{figure=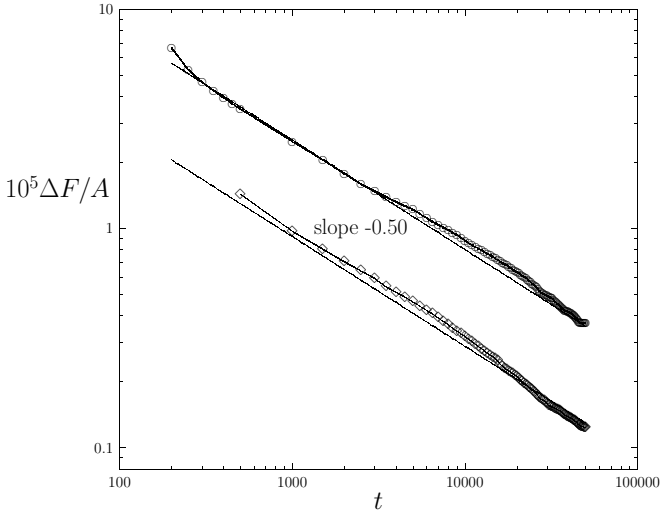,width=4.in}
\vspace{-5.5cm}
\caption{\label{ener} Relative Liapunov free-energy per unit area 
($A$ is the system area) as a function of time.
The legend is the same as in Figure \ref{gb}. 
Solid lines are guides to the eye.}
\end{figure}

We study the time evolution of the
defect density $\rho_d$, defined as the fraction of area occupied by the
black regions in Fig. \ref{t1200def}.
We perform three series of runs at small quenches, each satisfying 
$\epsilon= 1.9|r_c|$ for different choices of the 
parameters $\{c,\eta,r\}$  ($\epsilon=0.0372,\,0.0184,\,0.0345$ respectively). 
Figure \ref{gb} shows a summary of the data obtained. 
In each cases, the results are consistent with the law 
\begin{equation}\label{density}
\rho_d\sim t^{-1/2},
\end{equation}
which corresponds to a defect characteristic length scale growing 
as $t^{1/2}$. This result seems to be fairly independent of the 
angle $\theta$.

We next investigate the time evolution of the Liapunov functional $F$
given by Eq.(\ref{liap}). If $F_0$ denotes the value of $F$ for
a perfectly ordered system, then the quantity $\Delta F=F-F_0$ represents
the excess energy due to defects. The system free energy $F$ decreases like
the total length of grain boundaries.
From (\ref{density}), one should expect
\begin{equation}\label{energy}
F-F_0\sim t^{-1/2}.
\end{equation}
The numerical data plotted in Figure \ref{ener} for small values of 
$\epsilon$ are consistent with this scaling relation as well.

\begin{figure}

\vspace{-2.5cm}
\hspace{-2.0cm} \epsfig{figure=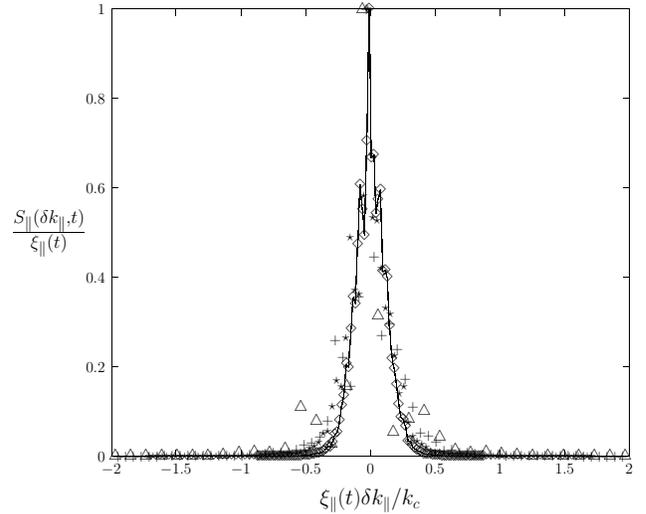,width=4.in}
\vspace{-5.5cm}
\caption{\label{scal} Structure factor $S_{\parallel}(\delta k_{\parallel},t)$ 
at four different times:
$t=5\, 10^2(\diamondsuit),\, 2\, 10^3(*),\, 5\, 10^3(+),\, 2\, 10^4(\Delta)$. 
At each time, the curve has been rescaled according to Eq.(\ref{scalSpara}), 
where $\xi_{\parallel}(t)$ is defined as $S(\delta k_{\parallel}=0,t)$. 
The parameters are $c=6,\, \eta=0.25,\, \epsilon=1.9|r_c|$. 
(Averages over 40 independent runs.)}
\end{figure}

The time evolution of many coarsening systems is self-similar: 
the large scale structure of successive 
configurations is statistically time invariant, provided that 
spatial variables are rescaled by a proper length. 
We thereby analyze the structure factor, defined as the Fourier 
transform of the two-point correlation function, 
$S(\vec{k})=\langle \psi(\vec{k},t)\psi(-\vec{k},t)\rangle$, 
the braces representing average over initial conditions.
$S(\vec{k})$ is maximum for any of the (four) wavevectors 
$\vec{k}_Z$ characterizing zig and zag rolls. At any given time,
we numerically observe that $S$ is maximal
for $\vec{k}_Z\approx\vec{k}_c$, given
by Eq.(\ref{oblique}). Therefore, the selected stripe periodicity 
and orientation in the polycrystalline structure are that of the 
marginal wavevectors $\vec{k}_c$ determined from linear stability analysis. 
A similar situation is encountered for the isotropic Swift-Hohenberg model, 
and is thought to be more generally a property of potential 
systems \cite{re:cross95a}. Near any peak $\vec{k}_Z$ of $S$,
we propose the following scaling ansatz
\begin{equation}\label{scalS}
S(\delta k_{\parallel},\delta k_{\bot},t)=\xi_{\parallel}(t)
\xi_{\bot}(t)\ f\left[\xi_{\parallel}(t)\delta k_{\parallel},
\xi_{\bot}(t)\delta k_{\bot}\right],
\end{equation}
where $\delta \vec{k}=\vec{k}-\vec{k}_Z=\delta k_{\parallel}\hat{k}_{\parallel}
+\delta k_{\bot}\hat{k}_{\bot}$,
with $\hat{k}_{\parallel}$ and $\hat{k}_{\bot}$ denoting the unit vectors
longitudinal and transverse to the wavevector $\vec{k}_{Z}$, respectively.
$f(x)$ is a scaling function, $\xi_{\parallel}(t)$ and $\xi_{\bot}(t)$ are
{\it a priori} two characteristic lengths describing grain growth in
the directions normal and parallel to the rolls, respectively.
Let us define $S_{\parallel}(\delta k_{\parallel},t)
\equiv\int_{-\infty}^{\infty}
\delta k_{\bot} S(\delta\vec{k},t)$, a symmetric relationship defining 
$S_{\bot}$. From (\ref{scalS}), one obtains the scaling ansatz
\begin{equation}\label{scalSpara}
S_{\parallel}(\delta k_{\parallel},t)=\xi_{\parallel}(t)
\ g\left[\xi_{\parallel}(t)\delta k_{\parallel}\right]\ .
\end{equation}

\begin{figure}

\vspace{-2.5cm}
\hspace{-2.5cm} \epsfig{figure=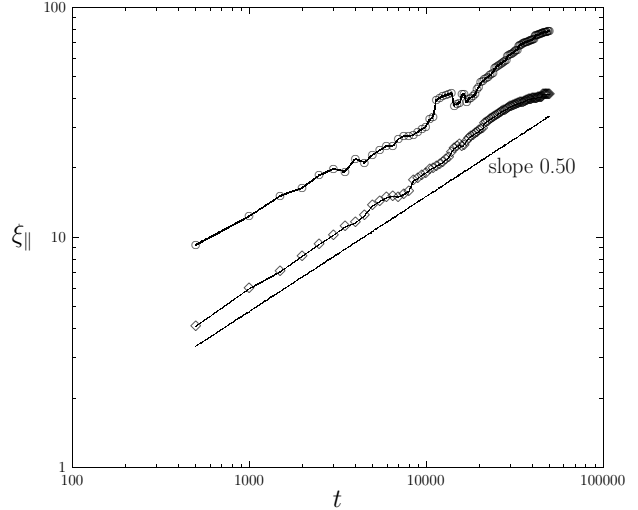,width=4.in}
\vspace{-5.5cm}
\caption{\label{struc} Length determined from the maximal intensity 
of the structure factor. See Figure \ref{gb} for legends. The
solid line is a guide to the eye.}
\end{figure}

Figure \ref{scal} displays $S_{\parallel}$ as a function
of $\delta k_{\parallel}$ at various times 
($t=5\, 10^2,\, 2\, 10^3,\, 5\, 10^3,\, 5\, 10^4$), for
$\{c=6,\, \eta=0.25,\, \epsilon=1.9|r_c|\}$. 
The scaling relation (\ref{scalSpara}) holds over nearly two decades,
despite of a slight widening at large times. 
The length $\xi_{\parallel}(t)$, taken from the maximum value of 
$S_{\parallel}$ is plotted as a function of time in Fig.\ref{struc}.
The results are in good agreement with
\begin{equation}\label{eq:xi}
\xi_{\parallel}(t)\sim t^{1/2}.
\end{equation}
We have not found $\xi_{\bot}(t)$, determined from $S_{\bot}$,
to be a convenient length scale to characterize coarsening.
$\xi_{\bot}$ is larger than $\xi_{\parallel}$ by
a factor varying between 5 (at short times) and 2 (at large times).
This could be due to some phase correlations of longer range than those 
associated with domain walls; these features were not investigated.

The above results are consistent with a coarsening process
driven by grain boundary surface tension and involving a single
characteristic length scale, like for the dynamics of Model A
for a non-conserved order parameter \cite{re:hohenberg77}. 
The law $R\sim t^{1/2}$ is also expected to describe the 
kinetics of grain growth in solid polycrystals \cite{re:mullins56}.
This situation differs markedly from isotropic stripes 
(where $R\sim t^{1/3}$ \cite{re:boyer01b,re:boyer02a,re:qian03}).
Note that the available experimental studies on anisotropic stripes have
reported much slower coarsening laws than Eq. (\ref{eq:xi}), namely
$t^{1/5}$ or $t^{1/4}$ \cite{re:purvis01,re:kamaga03}.

\section{Coarsening at larger quench depths}
\label{sec:pinning}

\subsection{Phenomenology of pinning forces in modulated phases}
\label{sec:pingen}

In solid crystals, for an isolated dislocation to glide from
one raw of atoms to the next one, there is a finite energy cost corresponding
to the raws that have to be compressed or dilated
during the move. The resulting elastic force, the Peierls stress,
tends to prevent the glide of dislocations \cite{re:peierls40,re:nabarro87}. 
Therefore, dislocations tend to be 
pinned in positions (periodically spaced) of minimum energy, and
motion takes place only if an external stress
larger than the critical Peierls stress is applied.
Remarkably, the defects present in systems that form
periodic patterns are also subjected to similar pinning forces.
Their origin is nonlinear in that case, and is due to the apparition
of \lq\lq non-adiabatic" terms in weakly nonlinear expansions. Studies on the 
Swift-Hohenberg model have shown that the laws of motion of grain boundaries 
(or dislocation arrays) involve short range, spatially periodic
pinning forces \cite{re:pomeau86,re:malomed90,re:boyer02a,re:boyer02b}. 
Either in crystals or in
patterns, the law of motion of a defect takes the general form
\begin{equation}\label{motionlaw}
\mu^{-1}v=\mu^{-1}dx/dt=f-p\cos(k_p x),
\end{equation}
where $v$ is the defect velocity, $x$ its position (for a grain boundary, 
the coordinate normal to the interface), $\mu$ a mobility,
$f$ an external force per unit length ({\it e.g}, 
the driving force for coarsening), 
and $p$ the magnitude of the pinning force, that oscillates with a periodicity
$2\pi/k_p$ proportional to the crystal (or base pattern) periodicity
$\lambda_c$.  Peierls-like pinning forces are usually much smaller than 
the other characteristic elastic forces (like the critical threshold shear
stress $f_{cr}$), and have 
the same general approximate expression, valid both for solids 
\cite{re:peierls40,re:nabarro87} and nonlinear patterns 
\cite{re:boyer02b,re:boyer02a}:
\begin{equation}\label{pgen}
p/f_{cr}\sim \exp[-aW/\lambda_c],
\end{equation}
where $W$ is the width of the defect (see further Fig. \ref{deepqlong}),
and $a$ a constant of order unity. For instance, grain boundaries separating 
domains of stripes have a width $W\sim \lambda_c/\sqrt{\epsilon}$
\cite{re:malomed90}. 
Therefore, close to the onset of the supercritical bifurcation 
($\epsilon\rightarrow 0^+$), $W$ becomes very large and the pinning
potential (\ref{pgen}) can be neglected. On the other hand, as the quench 
depth $\epsilon$ increases, $W$ decreases and pinning forces
can become large enough to affect qualitatively defect dynamics.

During a coarsening processes driven by surface tension, the average force
$f$ in Eq.(\ref{motionlaw}) is time dependent: $f\sim \gamma \kappa
\sim \gamma/R(t)$, 
with $\gamma$ and $\kappa$ the typical interface surface tension and 
curvature, respectively. At short times, domains have small sizes $R$, and
$f$ is large compared with $p$ in Eq. (\ref{motionlaw}). Defects move easily 
and the average grain size grows. As a result, the driving force $f$ decreases 
with time. At some point, $f$ may eventually become 
lower than the typical value of $p$ (which is time independent).  
In this case, boundaries
become pinned at one of the discrete stable positions $x_p$ such that 
$v=0$ in Eq.(\ref{motionlaw}). This situation is easy to observe numerically
for large quenches ($p$ large), where asymptotic patterns
remain only partially ordered (\lq\lq glassy"), with many immobile 
defects \cite{re:hou97,re:boyer02a}. On the other hand,
in the regime $\epsilon\rightarrow 0$, defect pinning is negligible
during the numerical time scales studied, and
coarsening dynamics is more likely to be self-similar and characterized
by well defined exponents.
 
For intermediate quenches, well before all grain boundaries of a system 
become pinned, the Peierls-like barriers are believed to slow down  
the ordering kinetics. 
At intermediate times, one may still be able to fit in some cases the
numerical results with power laws,
$R\sim t^{1/z^*}$. $z^*$ now represents an {\it effective} growth exponent, 
and $z^*\ge z|_{\epsilon\rightarrow 0}$.

\subsection{Pinning of anisotropic stripes}\label{sec:pinani}

To check whether the arguments presented above apply 
to anisotropic patterns as well, we have performed
calculations for deeper quenches than in Sec.\ref{lowq}.

\begin{figure}
\epsfig{figure=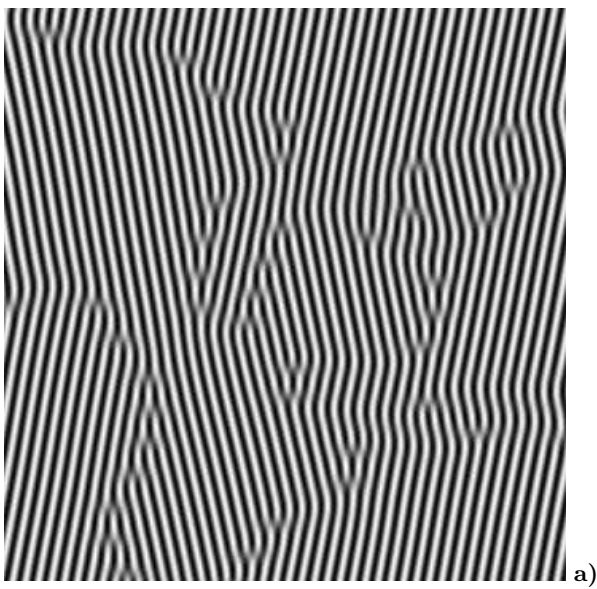,width=3.in,angle=90} {\bf a)}
%
\epsfig{figure=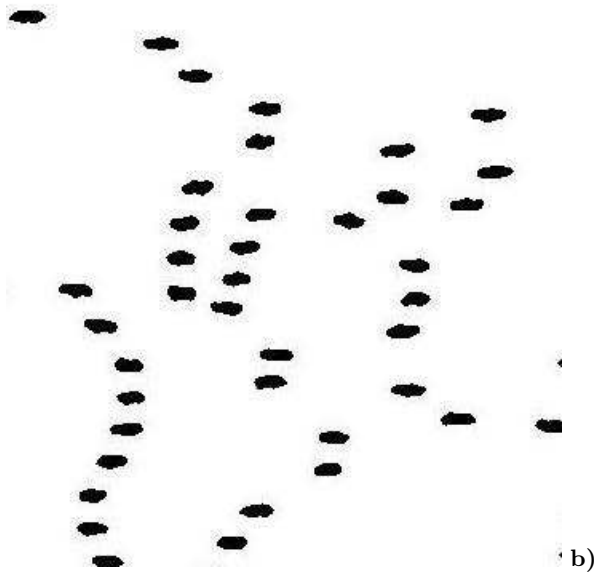,width=3.in,angle=90} {\bf b)}
\caption{\label{deepq} {\it a}) Local order parameter in gray scale, $c=12$,
$\eta=0.5$, $t=500$. The quench is \lq\lq moderate": 
$\epsilon=11|r_c|=0.215$.
The chevron boundaries are now straight and pinned.
{\it b}) Dislocation field of {\it a}) (same scale) obtained with the Fourier 
filtering procedure.}
\end{figure}

\begin{figure}
\epsfig{figure=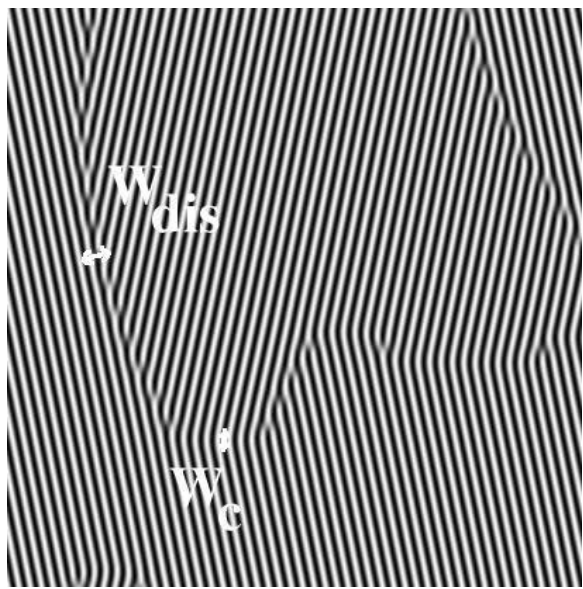,width=3.in}
\caption{\label{deepqlong} Same parameters as in Fig. \ref{deepq}, at
$t=50000$ (detail). The chevron boundaries and dislocations have
a width $W_c$ and $W_{dis}$ respectively.}
\end{figure}

Figure \ref{deepq}a shows in gray scale the field $\psi$ of a 
fraction of the system, obtained for a \lq\lq moderate" quench 
of depth $\epsilon=11|r_c|=0.215$, 
at time $t=500$ with $\{c=12,\,\eta=0.5\}$.
The two classes of defects previously mentioned, the
horizontal chevron grain boundaries and the dislocations,
can now be clearly distinguished. 
(The dislocation field is shown in Fig.\ref{deepq}b.)
The chevron boundaries are fairly straight. They remain practically 
immobile during the whole coarsening process, that is driven by
dislocation motion only.
This feature was observed in experiments as well \cite{re:purvis01,re:kamaga03}.
A detail of a large time configuration ($t=50000$) is shown in
figure \ref{deepqlong}. Like for shallow quenches, dislocations
tend to organize along string-like structures that are generally curved. 
We interpret the immobility of the chevron boundary as caused by strong
pinning forces. Given two domains of zig 
and zag rolls, the stable positions of a chevron 
boundary are imposed by the phase of the local order parameter, which 
does not change across the boundary when one follows a given roll. 
On the other hand, dislocations are much more mobile,
suggesting that their pinning potential $p$ is very low, and therefore
strongly anisotropic with respect to the grain boundary orientation.

In the following, we define the effective exponents associated to dislocations,
free-energy and structure factor, respectively:
\begin{equation}
\rho_{dis}\sim t^{-1/z^*_{dis}},\ \Delta F\sim t^{-1/z^*_F},\
\xi_{\parallel}(t)\sim t^{1/z^*_S},
\end{equation}
where $\rho_{dis}$ is the dislocation density, and is determined the same 
way as $\rho_d$ in Section \ref{lowq} (the fraction of black area of 
Fig.\ref{deepq}b) \cite{comment}. The other quantities have been defined in 
Section \ref{lowq}.

\begin{figure}

\vspace{-2.5cm}
\hspace{-2.5cm} \epsfig{figure=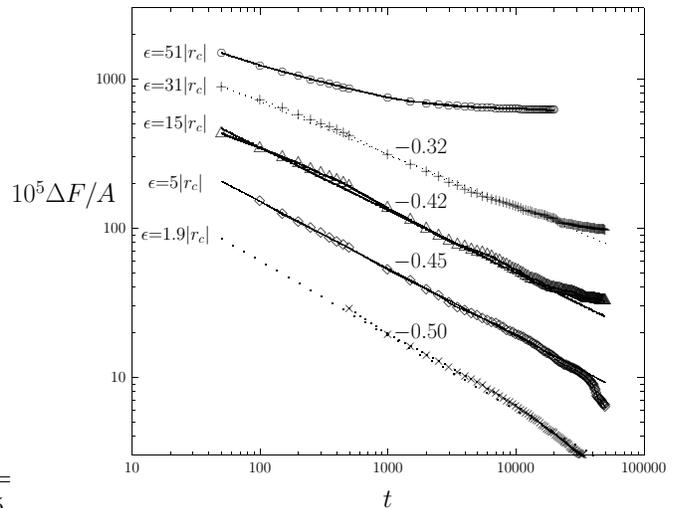,width=4.in}
\vspace{-5.5cm}
\caption{\label{pinning.en} Relative free energy per unit area 
as a function of time for
$c=12$, $\eta=0.5$ and various quench depths. From bottom to top:
$\epsilon=0.037;\, 0.098;\, 0.294;\, 0.607;\, 1$.}
\end{figure}

\begin{figure}

\vspace{-2.5cm}
\hspace{-2.5cm} \epsfig{figure=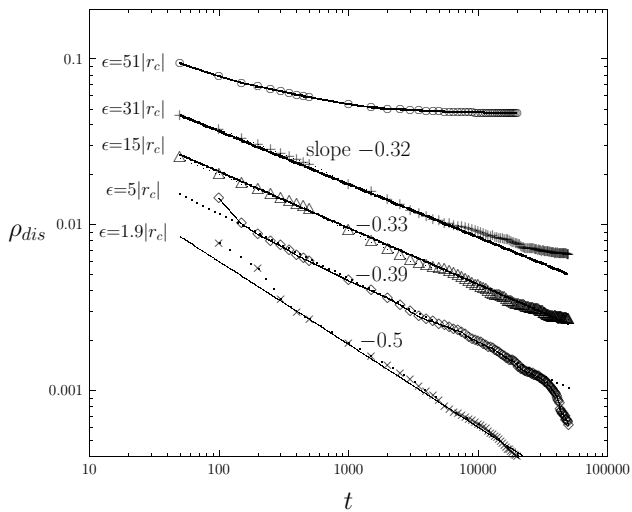,width=4.in}
\vspace{-5.5cm}
\caption{\label{pinning.dis} Dislocation density 
as a function of time. Same parameter as in Fig.\ref{pinning.en}.}
\end{figure}

We have plotted the time evolution of the relative free-energy per
unit area $\Delta F$ in Fig.\ref{pinning.en}, for different quench depths.
Similar curves are obtained for the dislocation density $\rho_{dis}$
(Fig.\ref{pinning.dis}), and the correlation length $\xi$ (not shown).
Provided that $\epsilon\le 35|r_c|$, the curves are still reasonably well 
fitted by power-laws during the first few decades considered in the
numerical calculations. At larger quenches, they rapidly saturates 
a finite values, indicating defect pinning. In these
cases, we define the effective exponent (arbitrarily) 
as given at time $t=1000$: 
\begin{equation}\label{defzeff}
z^*_{dis}=
-\left(\left. \frac{d\ln \rho_{dis}}{d\ln t}\right|_{t=1000}\right)^{-1}
\end{equation}
(and similar relations for $z^*_F$ and $z^*_S$.)

\begin{figure}

\vspace{-2.5cm}
\hspace{-2.0cm} \epsfig{figure=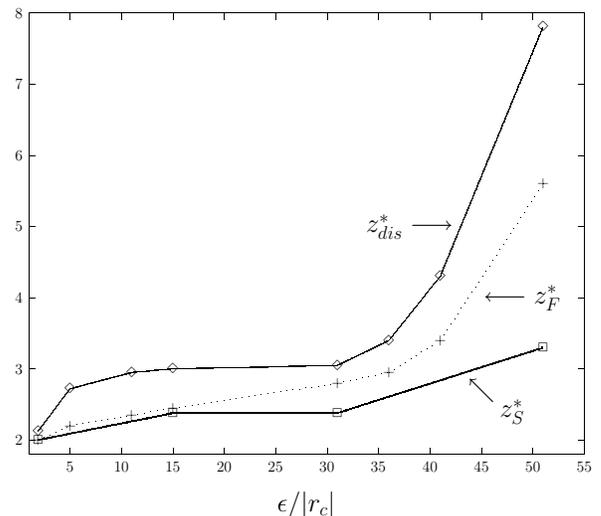,width=4.in}
\vspace{-5.5cm}
\caption{\label{zeff} Effective exponents for dislocations, energy, and 
structure factor as a function of the reduced quench depth $\epsilon/|r_c|$. 
The parameters in Eq.(\ref{pk}) are $c=12$, $\eta=0.5$.}
\end{figure}

As expected from the discussion of Section \ref{sec:pingen},
the ordering kinetics slow down noticeably as $\epsilon$ is increased.
All exponents $z^*_{dis}$, $z^*_F$ and $z^*_S$ increase with $\epsilon$.
Figure \ref{zeff}, displays the variations of the different
effective exponents as a function of $\epsilon$. 
The dislocation exponent $z^*_{dis}$ differs noticeably from
$z^*_{F}$ (and $z^*_S$): $z^*_{dis}>z^*_{F}>z^*_{S}$.
The behavior of $z^*_F$ is characterized by two regimes:
At moderate quenches, $z^*_F$ gradually departs from $z^*_F=2$ and
slowly increases with $\epsilon$ up to a value close to $3$.
For $\epsilon\ge35|r_c|$, $z^*_F$ then increases more sharply, the signature
of a sudden increase of pinning effects. A similar behavior (although
less pronounced) is observed for $z^*_S$.
The behavior of $z^*_{dis}$ with the quench depth is more abrupt.
The variations of $z^*_{dis}$ are quite important 
for small and large values of $\epsilon$.
The most striking feature is the presence of a fairly long plateau at 
intermediate quenches ($5|r_c|<\epsilon<35|r_c|$) where $z^*_{dis}$ remains 
practically constant, $z^*_{dis}\simeq 3$.
This result agrees with the experimental results 
of \cite{re:kamaga03}, where a law $t^{-1/3}$ was reported for the 
dislocation density. 

\subsection{Discussion}\label{sec:dis}

We sketch a possible interpretation of part of the above observations, 
based on the fact that pinning effects are strongly anisotropic. We saw that, 
away  from onset, chevron boundaries become totally pinned. 
Meanwhile, dislocations are mobile and may still have a very low pinning 
potential. From the standard relation for the width of a
grain boundary in stripe patterns, derived from weakly nonlinear 
analysis \cite{re:malomed90}, let us assume that
the width of chevron boundaries
($W_c$) and that of a roughly vertical dislocation array ($W_{dis}$, 
see Fig. \ref{deepqlong})
are given by $W_{c,dis}=\delta_{c,dis}\ \lambda/\sqrt{\epsilon}$, 
with $\delta_c$ and $\delta_{dis}$ two constants of order unity. 
From observations, let us assume that dislocations have a larger
width than chevrons: $W_{dis}>W_c$, {\it i.e.} $\delta_{dis}>\delta_c$.

From the general relation (\ref{pgen}), the pinning potentials have an 
non-analytical behavior at low $\epsilon$, and saturate at large $\epsilon$
($p\sim\exp[-{\rm cst}/\sqrt{\epsilon}])$. If $\delta_{dis}>\delta_c$, 
the pinning potentials $p_{c}$ and $p_{dis}$ for chevrons and 
dislocations respectively, are such that 
$p_{dis}\ll p_{c}\ll 1$ if $\epsilon\ll 1$.
A moderate increase of $\epsilon$ can cause a rapid increase of $p_c$
up to its saturation value, while $p_{dis}$ may still remain very low. 
This situation can happen in the range of quench depths defined by 
$(a\delta_{c})^2<\epsilon< (a\delta_{dis})^2$ 
($a$ is introduced in Eq.(\ref{pgen})), provided that this range is 
sufficiently broad. Hence,
one would expect the coarsening dynamics to be
relatively insensitive to the value of $\epsilon$ in that range, 
as observed numerically for $5|r_c|<\epsilon<35|r_c|$ in Fig. \ref{zeff}.
Further increase of $\epsilon$ eventually produces dislocation pinning,
and a general slowing down of the system must occur: it is illustrated by 
the inflection of the effective exponents past $35|r_c|$.

Let us assume next that the system is described by
two characteristic length scales, $L_{dis}$ and $L_c$, representing
the linear extent of a grain along the $y$ and $x$ directions, respectively.
Following the arguments of \cite{re:kamaga03}, the dislocation density can 
be written as $\rho_{dis}\sim L_{dis}/(L_{dis}L_c)=L_c^{-1}$.
In the moderate quench regime ($5|r_c|<\epsilon<35|r_c|$),
$L_{c}\sim t^{1/3}$. In this regime, the results on the energy and the
structure factor (Fig.\ref{zeff}) suggest that the other length scale,
$L_{dis}$, grows faster than $t^{1/3}$ ($2\le z^*\le 3$). Therefore, as 
time goes, defected regions tend to be more composed of dislocations than
chevron boundaries ($L_{dis}>L_c$). Unfortunately,
this finding disagrees with the
experimental results on electroconvection, 
where the opposite behavior was found: a very slow growth
law for $L_{dis}$ ($\sim t^{1/5}$), as well as a similar law
for a correlation length, 
was reported in \cite{re:kamaga03}. Hence, the 
experimental grains are elongated along the $x$ direction at 
late stages \cite{re:kamaga04}.

\section{Conclusions}\label{sec:concl}

We have presented evidence that the coarsening of smectic patterns, as 
given by a potential anisotropic Swift-Hohenberg equation, is characterized 
by a $t^{1/2}$ law close to onset.
This law has not been observed experimentally so far in 
electroconvection of nematic liquid crystals,
and may correspond to a regime difficult to reach.
For larger quench depths, the phase ordering kinetics is affected
by pinning effects that strongly depend on grain boundary orientations. 
A particular regime is observed numerically for a fairly wide interval 
of moderate quenches: chevron boundaries get pinned, and grain growth still 
takes place via mobile arrays of dislocations. A similar behavior was 
observed in experiments \cite{re:purvis01,re:kamaga03}.
In this regime, the dislocation density behaves as $t^{-1/3}$, the
same decay rate as found experimentally \cite{re:kamaga03}. 
Our results suggest that the characteristic length of a
dislocation array grows faster than that of a chevron boundary. This feature 
disagrees with the experimental findings, though, and may points out a 
limitation of the present model. Nonpotential effects, that have been
neglected here, probably play an important role in this system.

\begin{acknowledgments}
We thank M. Dennin and C. Kamaga for fruitful discussions and for 
communicating us unpublished experimental results. This work was supported
by the Consejo Nacional de Ciencia y Tecnolog\'\i a (CONACYT, Mexico) 
Grant number 40867-F.
\end{acknowledgments}

\bibliographystyle{prsty}

\begin{thebibliography}{10}
\bibitem{comment}
The filtering procedure is such that, 
at as soon as $\epsilon\ge 5|r_c|$, dislocations only are emphasized,
while the extended chevron boundaries, previously 
visible in Fig.\ref{t1200def} at shallow quenches, 
no longer appear.  Therefore, in the moderate-large quench regime, 
the defect density 
determined from filtering has to be identified with $\rho_{dis}$ only, and 
no longer with the total amount of defects of the system $\rho_d$
(dislocations + chevron boundaries).
\end{thebibliography}

\end{document}